\documentclass[conference]{IEEEtran}
\IEEEoverridecommandlockouts
\usepackage{cite}
\usepackage{authblk}
\usepackage{subcaption}
\usepackage{amsmath,amssymb,amsfonts}
\usepackage{graphicx}
\usepackage{textcomp}
\usepackage{xcolor}
\usepackage[english]{babel}
\usepackage[utf8]{inputenc}
\usepackage{algorithm}
\usepackage[noend]{algpseudocode}

\def\BibTeX{{\rm B\kern-.05em{\sc i\kern-.025em b}\kern-.08em
    T\kern-.1667em\lower.7ex\hbox{E}\kern-.125emX}}
\begin{document}

\title{Memory Leak Detection Algorithms in the Cloud-based Infrastructure 
}

\author[1]{Anshul Jindal}
\author[2]{Paul Staab}
\author[1]{Pooja Kulkarni}
\author[2]{Jorge Cardoso}
\author[1]{Michael Gerndt}
\author[1]{Vladimir Podolskiy}
\affil[1]{\textit{Chair of Computer Architecture and Parallel Systems, TU Munich, Garching, Germany }}
\affil[1]{\textit {{\{anshul.jindal, pooja.kulkarni\}@tum.de, gerndt@in.tum.de, v.podolskiy@tum.de}}}
\affil[2]{\textit{Huawei Munich Research Center, Huawei Technologies Munich, Germany}}
\affil[2]{\textit {{\{paul.staab, jorge.cardoso\}@huawei.com}}}
\renewcommand\Authands{ and }

\maketitle

\begin{abstract}

A memory leak in an application deployed on the cloud can affect the availability and reliability of the application. Therefore, to identify and ultimately resolve it quickly is highly important. However,  in the production environment running on the cloud, memory leak detection is a challenge without the knowledge of the application or its internal object allocation details. 

This paper addresses this \textit{challenge of detection of memory leaks in cloud-based infrastructure without having any internal knowledge} by introducing two novel machine learning based algorithms: Linear Backward Regression (LBR) and Precog and, their two variants: Linear Backward Regression with Change Points Detection (LBRCPD) and Precog with Maximum Filteration (PrecogMF). These algorithms only use one metric i.e the system's memory utilization on which the application is deployed for detection of a memory leak. The developed algorithm's accuracy was tested on 60 virtual machines manually labeled memory utilization data provided by Huawei and it was found that the proposed PrecogMF algorithm achieves the highest accuracy score of 85\%. The same algorithm also achieves this by decreasing the overall compute time by 80\% when compared to LBR's compute time.

The paper also presents the different memory leak patterns found in the various memory leak applications and are further classified into different classes based on their visual representation.

\end{abstract}

\begin{IEEEkeywords}
memory leak, online memory leak detection, memory leak patterns, cloud, linear regression
\end{IEEEkeywords}

\section{Introduction}

Cloud computing is widely used in the industries for its capability to provide cheap and on-demand access to compute and storage resources. In the cloud  physical servers located at different data centers resources are split among the VMs hosted on it and distributed across the users~\cite{7570950}. Users can then deploy their applications on these VMs with only the required resources. This allows the efficient usage of the physical hardware and reducing the overall cost. However, with all the advantages of cloud computing there exists the drawback of detecting a fault or an error in an application or in a VM efficiently due to the layered architecture~\cite{7416355, 7977362}. A small fault somewhere in the system can impact the performance of the application. 

An application when deployed on a VM usually requires different system resources such as memory, CPU and network for the completion of a task. If an application is mostly using the memory for the processing of the tasks then  this application is called a memory-intensive application~\cite{6799629}. It is the responsibility of the application to release the system resources when they are no longer needed. When such an application fails to release the memory resources, a \textbf{memory leak} is said to occur in the application~\cite{Xie:2005:CPM:1095430.1081728}. Memory leak issues in the application can cause continuous blocking of the VM's resources which may in turn result in slower response times or application failure. In software industry, memory leaks are treated with utmost seriousness and priority as the impact of a memory leak could be catastrophic to the whole system. In the development environment, these issues are rather easily detectable with the help of static source code analysis tools or by analyzing the heap dumps. But in the production environment running on the cloud, memory leak detection is a challenge and it only gets detected when there is an abnormality in the run time, abnormal usage of the system resources, crash of the application or restart of the VM. Then the resolution of such an issue is done at the cost of compromising the availability and reliability of the application. Therefore it is necessary to monitor every application for memory leak and have an automatic detection mechanism for memory leak before it actually occurs. However, it is a challenge to detect memory leak of an application running on a VM in the cloud without the knowledge of the programming language of the application, nor the knowledge of source code nor the low level details such as allocation times of objects, object staleness, or the object references~\cite{Sor2011ASA}. Therefore, this challenge is addressed in this paper  \textit{ by only using the VM's memory utilization as the main metric and devising four different algorithms to detect memory leak}. Due to the low down time requirements for the applications running on the cloud, detection of issues and their resolutions is to be done as quickly as possible. Therefore, this point is also kept into consideration while creating the memory leak detection algorithms.

If an application has a memory leak, usually the memory usage of the VM on which it is running increases steadily. It continues to do so until all the available memory of the system is exhausted. This usually causes the application attempting to allocate the memory to terminate itself. Thus, usually a memory leak behaviour exhibits a linearly increasing or "sawtooth" memory utilization pattern. However, to verify such an argument \textit{different memory leak application's memory utilization data is collected and studied in this work . These patterns are then classified into different classes based on their visual representation}. 

The main contributions of this paper are summarized as follows:
\begin{itemize}
 \item We propose two novel machine learning based algorithms: Linear Backward Regression (LBR) and Precog and, their two variants: Linear Backward Regression with Change Points Detection (LBRCPD) and Precog with Maximum Filteration (PrecogMF) for accurate and efficient detection of memory leaks by only using the VM's memory utilization as the main metric. It was found that, the Proposed PrecogMF achieves the best accuracy score of 85\% on the evaluated dataset provided by Huawei. 
 \item We also presents different memory leak patterns found after conducting experiment on 6 different memory leaking applications and classify them into different classes based on their visual representations. 
 \end{itemize}
 
The rest of this paper is composed as follows. 
Section 2 discusses background knowledge required for this paper in brief. Section 3 studies the related works. Section 4  presents different memory leak patterns observed. Section 5 describes the overall problem statement and the followed approach to solve it. Section 6 provides the design and details of the different memory leak algorithms developed. Section 7 provides experimental configuration details along with the hyper-parameters values used and, also showcase results of the conducted analysis. Section 8 summarizes the discussion of the results and lastly, Section 9 concludes the paper.

\section{Background}
In this section, we present the basic background knowledge required for this paper.
\subsection{Memory management in garbage collected programming languages }
Automatic garbage collection is one of the most important aspects of modern day programming languages, which simplifies the life of application developers taking away the burden of manual memory management. Programming languages such as \textbf{C} and \textbf{C++}, memory related faults occur due one of the following reasons \textit{1. Dangling pointers} - de-referencing pointers to objects that the program previously freed, \textit{2 lost pointers} – losing all pointers to objects that the program neglects to free, and \textit{(3) unnecessary references} – keeping pointers to objects the program never uses again. In garbage collected programming languages, GC algorithms are designed to automatically deal with \textit{1 and 2}. However, keeping a reference to an unused object can not be tracked by garbage collection and program starts leaking memory. Thus, a memory leak in a garbage-collected language occurs when a program continue to keep the references to objects that are no longer needed, preventing the garbage collector from reclaiming space. Best in such cases, the object is not a growing instance, and results in constant amount of memory being leaked. But in worst cases, a growing object with unused parts will cause the program/applications to run out of memory and crash~\cite{Jump:2007:CDM:1190216.1190224}. These type of memory fault defects are very difficult to debug and fix given the critical nature of the defect as the ultimate object that crashes the application may not necessarily be of leaking nature. 

\subsection{Linear Regression}
Linear regression is an approach to model the relationship between one or more variables~\cite{montgomery2012introduction}. This is done to ensure generalization and give the model the ability to predict the outputs from the new input values~\cite{freedman_2009}. The case where mapping of one independent variable is done to one dependent variable is called univariate linear regression or simple linear regression~\cite{freedman_2009}. Applying simple linear regression means mapping the input variable to the output variable i.e fitting a linear line onto the input data points. In machine learning convention this is denoted by equation~\eqref{eq:line}.  
\begin{equation}
h(x) = \theta_0 + \theta_1.x\label{eq:line}
\end{equation}
where $\theta_0$ and $\theta_1$ are weights, $x$ is the input variable and $h(x)$ is the label or the output value. 

When the simple linear regression is applied, it tries to find the weights $\theta_0$ and $\theta_1$ that can give the best-fitted line for the input data~\cite{Mihaescu2011ClassificationOL}. This best fitted line is determined to keep the error loss minimum. This error loss in linear regression is calculated based on a cost function ~\eqref{eq:cost_function}. 
\begin{equation}
J(\theta_0, \theta_1) = \frac{1}{2n}\sum\limits_{i=1}^n(h(x^{(i)}) - y^{(i)})^2\label{eq:cost_function}
\end{equation}
Where $J(\theta_0, \theta_1)$ refers to the total cost of the model with weights $\theta_0$ and $\theta_1$. $h(x^{(i)})$ refers to the model’s prediction at input $x^{(i)}$. $y^{(i)}$ is the actual output value at index $i$ and, $n$ is the total number of data points in the dataset.
The cost function basically gets the distance for example euclidean distance between the actual y-value $y^{(i)}$ and the model's predicted y-value $h(x^{(i)})$, then squaring this distance for every point and summing up and dividing it by the number of data points in the dataset to get the average cost. Different algorithms can be used for learning the the best weights $\theta_0$ and $\theta_1$ that minimize the cost~\cite{weisberg2005applied}. Gradient Descent is one such algorithm which uses the two equations~\eqref{eq:theta0} and~\eqref{eq:theta1} until convergence is reached. 
\begin{equation}
\theta_0 := \theta_0 - \alpha \frac{\partial J(\theta_0, \theta_1)}{\partial  \theta_0}\label{eq:theta0}
\end{equation}
\begin{equation}
 \theta_1 := \theta_1 - \alpha\frac{\partial J(\theta_0, \theta_1)}{\partial  \theta_1}\label{eq:theta1}
\end{equation}
where $\alpha$ is the learning rate and $\frac{\partial }{\partial  \theta_0}$ and $\frac{\partial}{\partial  \theta_1} $ are partial derivatives of $J(\theta_0, \theta_1)$ with respect to $\theta_0$ and $\theta_1$
Running the gradient descent learning algorithm on the model, and through the costs obtained at every step, the model will converge to a minimum cost. The weights that led to that minimum cost are the final values for the best fitted-line on the input data.  In the context of the current work, we use linear regression to find the best-fitted trend line on the input time-series data.

\section{Related Work}
Memory leak detection has been studied over the years and several solutions have been proposed. Sor et al. reviewed different memory leak detection approaches based on their implementation complexity, measured metrics, and intrusiveness and a classification taxonomy was proposed~\cite{Sor2014MemoryLD}. The classification taxonomy broadly divided the detection algorithms into \textit{(1) Online detection, (2) Offline detection and (3) Hybrid detection}. The \textit{online detection} category uses either staleness measure of the allocated objects or their growth analysis. \textit{Offline detection} category includes the algorithms that make use of captured states i.e heap dumps or use a visualization mechanism to manually detect memory leaks or use static source code analysis. \textit{Hybrid detection} category methods combine the features offered by online and offline methods to detect memory leaks. Our work falls in the category of online detection therefore, we now restrict our discussion to the approaches related to the online detection category only. 

Based on the staleness measure of allocated objects, Rudaf et al. proposed "LeakSpot" for detecting memory leaks in web applications~\cite{Rudafshani:2017:LDD:3035064.3035070}. It locates JavaScript allocation and reference sites that produce and retain increasing numbers of objects over time and uses staleness as a heuristic to identify memory leaks. Vladimir \v{S}or et al. proposes a statistical metric called \textit{genCount} for memory leak detection in Java applications~\cite{Sor2015MemoryLD}. It uses the number of different generations of the objects grouped by their allocation sites, to abstract the object staleness - an important attribute indicating a memory leak.  Vilk et al. proposed a browser leak debugger for automatically debugging memory leaks in web applications called as "BLeak"~\cite{Vilk:2018:BAD:3192366.3192376}. It collects heap snapshots and analyzes these snapshots over time to identify and rank leaks. BLeak targets application source code when detecting memory leaks. 

Based on the growth analysis objects, Jump et al. proposes "Cork" which finds the growth of heap data structure via a directed graph  \textit{Type Points-From Graph} - TPFG, a data structure which describes an object and its outgoing reference~\cite{Jump:2007:CDM:1190216.1190224}. To find memory leaks, TPFG's growth is analyzed over time in terms of growing types such as a list. FindLeaks proposed by Chen et al. tracks object creation and destruction and if more objects are created than destroyed per class then the memory leak is found~\cite{4291098}. Nick Mitchell and Gary Sevitsky proposed "LeakBot", which looks for heap size growth patterns in the heap graphs of Java applications to find memory leaks ~\cite{10.1007/978-3-540-45070-2_16}. "LEAKPOINT" proposed by Clause et al. uses dynamic tainting to track heap memory pointers and further analyze it to detect memory leaks~\cite{6062055}.  

Most of the online detection algorithms that are proposed focus either on the programming language of the running application or on garbage collection strategies or the internals of the application based on the object's allocation, references, and deallocation. To the best of our knowledge, there is no previous work that solely focuses on the detection of memory leaks using just the system's memory utilization data on which application is deployed. The work in this paper, therefore, focuses on the detection of a memory leak pattern irrespective of the programming language of the application or the knowledge of application's source code or the low-level details such as allocation times of objects, object staleness, or the object references. 
\section{Memory Leak Patterns}
\subsection{Testbed Infrastructure}
For understanding the memory leak patterns, we created a monitoring pipeline to collect the resources usage statistics of a deployed application with the following tools: 
\subsubsection{Prometheus}
Prometheus is a de facto industry standard monitoring solution. It records real-time metrics in a time series database  with flexible queries and real-time alerting~\cite{prometheus}. A typical monitoring platform with Prometheus is composed of multiple tools: exporters to export host's metrics, Prometheus to centralize and store the metrics and alertmanager to trigger alerts based on those metrics.  Node exporter is one of the exporter built for collecting system metrics from the host machine and exporting them to Prometheus, e.g. CPU, memory and disk usage~\cite{nodeexporter}. cAdvisor is an exporter built for collecting metrics from inside the Docker containers. 

\subsubsection{InfluxDB}
InfluxDB is an open-source time series database. It is optimized for fast, high-availability storage and retrieval of time series data. It has no external dependencies and provides a SQL-like query language for querying the data. It has  built-in time-centric functions for querying a data structure composed of measurements, series, and points. Each point consists of several key-value pairs called the fieldset and a timestamp. When grouped together by a set of key-value pairs called the tagset, these define a series. Finally, series are grouped together by a string identifier to form a measurement.

\begin{figure}[htbp]
\centerline{\includegraphics[width=0.9\linewidth]{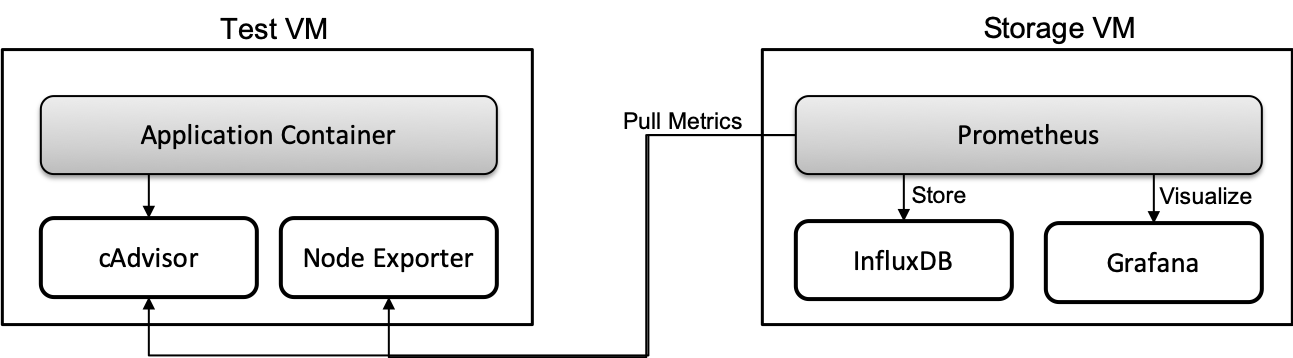}}
\caption{Overall workflow for collection of the VM memory utilization when an application is deployed on it. }
\label{overall_data_collection_method}
\end{figure}

The overall test environment infrastructure is shown in Fig.~\ref{overall_data_collection_method} which consists of two VMs : Test VM and Storage VM. Test VM runs the application container for which the data is to be collected along with the metrics collection tools:  cAdvisor for container metrics and Node exporter for node resources metrics collection. Storage VM runs the Prometheus tool which pulls the different metrics values from the cAdvisor and Node Exporter. These metrics values are saved in InfluxDB for future use and are visualized in Grafana for monitoring.  

\begin{figure*}[htbp] 
\begin{subfigure}{.33\textwidth}
  \centering
  \includegraphics[width=0.97\linewidth]{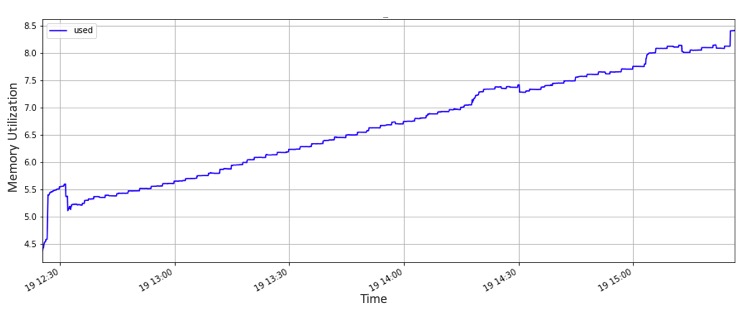}
  \captionof{figure}{A java application with memory leak}
  \label{fig:java_memory_leak}
\end{subfigure}%
\begin{subfigure}{.33\textwidth}
  \centering
  \includegraphics[width=0.97\linewidth]{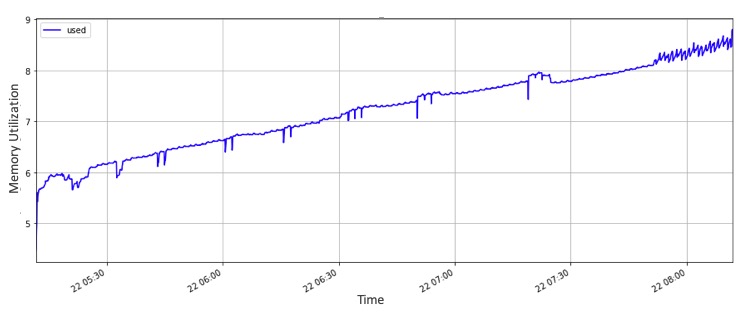}
  \captionof{figure}{A Node.js application with memory leak}
  \label{fig:node_js_memory_leak}  
\end{subfigure}
\begin{subfigure}{.33\textwidth}
  \centering
  \includegraphics[width=0.97\linewidth]{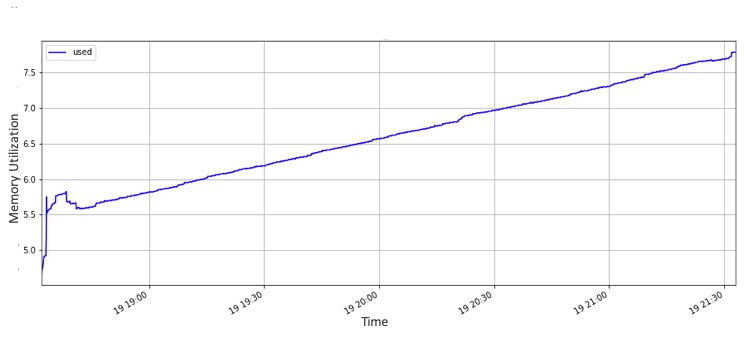}
  \captionof{figure}{A Python application with memory leak}
  \label{fig:python_mem_leak}  
\end{subfigure}
\vspace{-2mm}
 \caption{Linearly increasing memory leak pattern}
  \label{linear_increase_memory_leak_pattern}
  \vspace{-6mm}
\end{figure*}

\subsection{Found memory leak patterns}
We collected the memory utilization data for 6 different open source applications written in different programming languages containing memory leaks after containerizing them using Docker. The containerized version of the applications along with the link to the original versions can be found at this GitHub repository~\cite{application_studied}. The collected memory utilization data is classified into three patterns as explained below along with the applications detail. 
\subsubsection{Linearly increasing pattern}
This pattern represents the cases where the memory utilization is increasing continuously with some noise in between until the system's memory is exhausted. Fig.~\ref{linear_increase_memory_leak_pattern} shows the 3 application's memory usage pattern classified into this case. These three applications are:
    \begin{itemize}
    \item Java application: This application continuously creates an object of size 4MB of a class every second but does not free it up. As a result we can see in the Fig.~\ref{fig:java_memory_leak} the steady increase in memory utilization for it.
    \item Node.js application: This application appends the logs into a global variable every time it receives a request at its root endpoint, thus, it keeps on using the system's memory with every request as shown in the Fig.~\ref{fig:node_js_memory_leak}.
    \item Python application: This application appends a constant value into a global variable every second but does not free it up. As a result we can see in the Fig.~\ref{fig:python_mem_leak} the steady increase in memory utilization for it.
    \end{itemize}
\subsubsection{Random pattern}
\begin{figure}[htbp]
 \vspace{-4mm}
  \centering
  \includegraphics[width=0.45\textwidth]{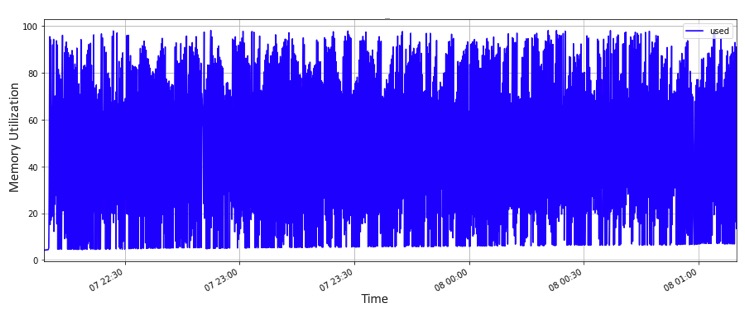}
    \caption{Random memory leak pattern}
    \label{random_memory_leak_pattern} 
     \vspace{-2mm}
\end{figure}
This pattern represents the cases where the memory utilization is random and hence unpredictable to find the leaking point for these applications. Fig.~\ref{random_memory_leak_pattern} shows an application's memory usage pattern classified into this case. This application is written in Java and creates a new instance of Java Inflater class continuously and does not properly end it.  Inflater class is used for general purpose decompression of the given data and if it is not properly ended, the uncompressed data will not be discarded and will be allocated heap memory. If the object is continuously created and not ended it will use more heap memory and  will result in memory leak. For detecting a memory leak in such a case other attributes values are required and hence not considered as part of this work.

\subsubsection{Saw-tooth pattern}
\begin{figure*}[htbp] 
\begin{subfigure}{.5\textwidth}
  \centering
  \includegraphics[width=0.94\linewidth]{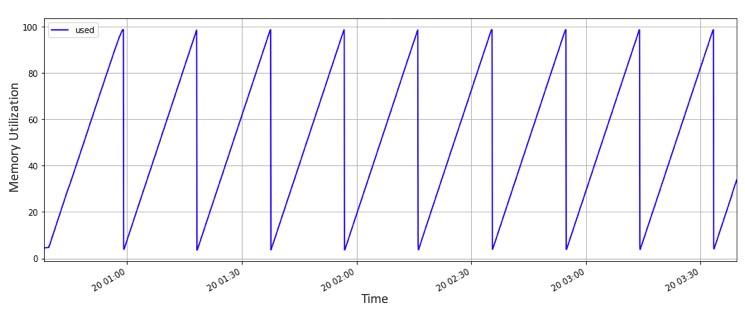}
  \vspace{-2mm}
  \captionof{figure}{Python application}
  \label{fig:saw_tooth_python}
\end{subfigure}%
\begin{subfigure}{.5\textwidth}
  \centering
  \includegraphics[width=0.94\linewidth]{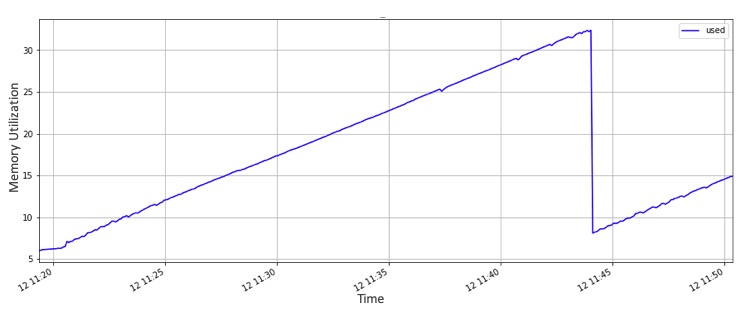}
    \vspace{-2mm}
  \captionof{figure}{Node.js application}
  \label{fig:saw_tooth_nodejs}  
\end{subfigure}
\vspace{-2mm}
 \caption{Saw-tooth memory leak pattern}
  \label{saw_tooth_memory_leak_pattern}
  \vspace{-4mm}
\end{figure*}
This pattern can be considered as the bigger version of the linearly increasing pattern, where the memory utilization of an application initially increases continuously until it reaches the maximum available point. After which if the application tries to allocate more memory, it either results in the crash or restart of the application as shown for example applications in Fig.~\ref{saw_tooth_memory_leak_pattern}. As part of this pattern two applications are studied: 
 \begin{itemize}
    \item Python application: This application continuously creates an object of a class every 0.1 second and does not free it up. As a result we can see a rapid memory growth for this application in the beginning until it reaches the maximum point after which the container restarts and again starts from 0 to the maximum point as shown in the Fig.~\ref{fig:saw_tooth_python}. This process continues until the application is stopped. 
    
    \item A Node.js application: This application appends a series of information into a global variable every second and thus keeps on using the system's memory  every second until either the memory is cleaned or the process is restarted.  Here the process is restarted every 25 seconds, that is why as shown in the  Fig.~\ref{fig:saw_tooth_nodejs} there is a continuous increase in the memory for initial 25 seconds and then it drops as the process is restarted and once it restarts it again continues to use the memory showcasing the saw-tooth pattern.
    \end{itemize}

The above experiment was only conducted to understand the memory usage patterns of a memory leak application and further use that understanding to develop memory leak detection logic. The dataset generated from the above applications cannot be used for the evaluation of the developed memory leak detection algorithms (discussed in the next sections), as these applications predominately only contain memory leaks (which is not always the case in real-world applications) and hence it would be wrong to evaluate the algorithms on them. Therefore, the dataset provided by our partner industry Huawei was used for evaluation. 

\section{Methodology for Memory Leak Detection}
In this section, we present the problem statement of memory leak detection and introduce our overall approach workflow for solving it. 
\subsection{Problem Statement}
A time series data contains successive observations which are usually collected at equal-space timestamps. In our study, we focus on
a virtual machine data from the Cloud, defined as $x = \{x_1, x_2, ..., x_N\}$, where $N$ is the length of $x$ and an observation $x_t \in R$ is the percentage utilization of a resource (for example memory or disk usage) of a virtual machine at time $t$ ($t \leq N$). A time series window of length $T$ referred by $x_{t - T:t}$ ($\in R_{(T +1}$) is used to denote a sequence of observations $\{x_{t - T}, x_{t - T+1}, ..., x_t\}$ from time ${t - T}$ to $t$.

\begin{figure}[htbp]
\centerline{\includegraphics[width=1\linewidth]{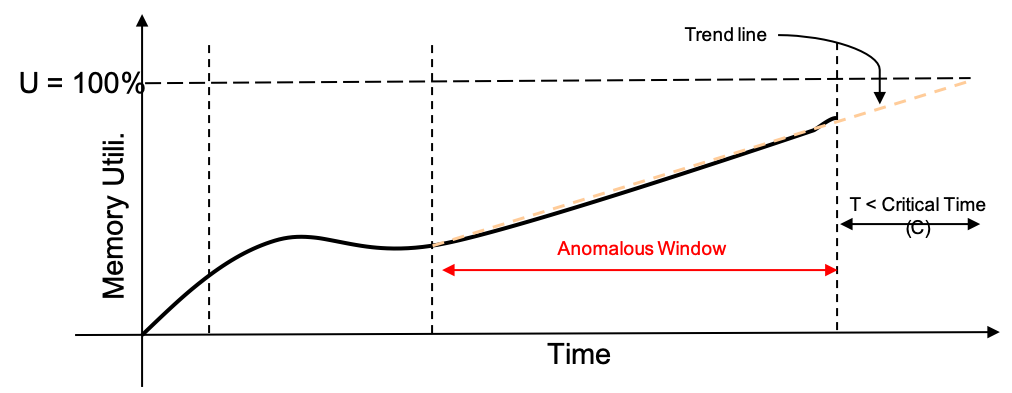}}
\caption{Example memory utilization of a memory leaking VM with the marked anomalous window.}
\label{fig_mem_leak_detection_problem}
\end{figure}

The current work focuses on the memory leak detection, therefore $x = \{x_1, x_2, ..., x_N\}$ represents the memory utilization observations of the VM. 
The objective of this work is to determine whether or not there is a memory leak on a VM such that an observation $x_t$ at time $t$ reaches the threshold $U$ ($U \leq 100$) memory utilization following a trend in the defined critical time $C$. 

Our approach returns an anomalous window for a VM consisting of a sequence of observations $x_{t - T:t}$ such that these observations after following a certain trend will reach the threshold $U$ memory utilization at time $t+M$ where $M \leq C$. The same approach can be applied for multiple VMs as well.  Fig.~\ref{fig_mem_leak_detection_problem} shows the memory utilization of an example memory leaking VM with the marked anomalous window. It shows that the memory utilization of the VM will reach the defined threshold ($U = 100\%$) within the defined critical time $C$ by following a linearly increasing trend (shown by the trend line) from the observations in the anomalous window. Therefore, this VM is regarded as a memory leaking VM.

\subsection{Appraoch}
 \begin{figure}[htbp]
\centerline{\includegraphics[width=0.9\linewidth]{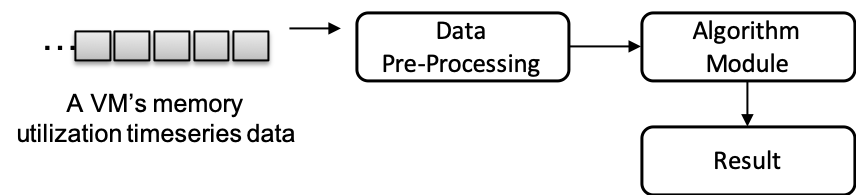}}
\caption{Overall followed approach workflow. }
\label{overall_followed_method}
 \vspace{-4mm}
\end{figure}
Fig.~\ref{overall_followed_method} displays the overall workflow of the followed approach. The entire system largely consists of two modules: \textit{Data Pre-processing} and \textit{Algorithm module}. In the current work, we have developed multiple algorithms and these algorithms are plugged in place of the \textit{Algorithm} module. Each algorithm's design and implementation are different and are explained in the further sections. \textit{Data Pre-processing} is a module common to all the algorithms. The procedure starts by collecting the memory utilization data of a VM and passing it to \textit{Data Pre-processing} module, where the dataset is first transformed by resampling  the number of observations to one every defined resampling time resolution and then the time series data is median smoothed over the specified smoothing window. For time series modeling, historical values are useful for understanding current data. Therefore, a sequence of observations $x_{t - L:t}$ instead of just $x_t$ belonging to a VM passed from \textit{Data Pre-processing} module is used by the individual algorithm module for a potential memory leak detection. If a window is found to be within the algorithm's criteria of memory leak then it will be reported as anomalous, otherwise, it is normal. Each algorithm has its criteria, design, and parameters for the discovery of a memory leak and is explained in the next section.

\section{Memory Leak Detection Algorithms}
In this section, we present in detail the different memory leak detection algorithms developed with respect to the overall approach introduced in the last section along with their design details.
%In addition, we provide also details about the hyper-parameters used and introduces new ones specific to the algorithm.
\subsection{Linear Backward Regression (LBR)}
 \begin{figure}[htbp]
 \vspace{-4mm}
\centerline{\includegraphics[width=1\linewidth]{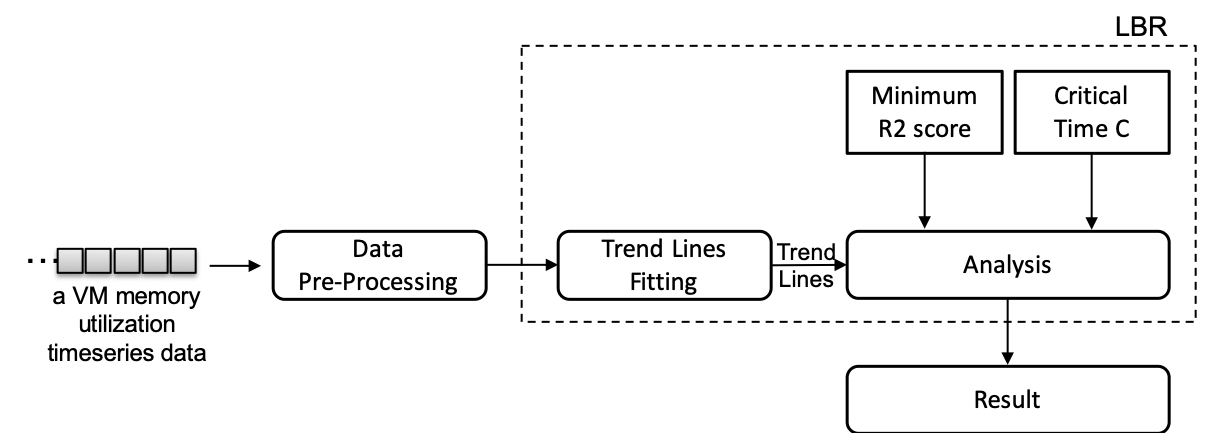}}
\caption{Overall workflow of the LBR algorithm. }
\label{fig_overall_lbr}
\end{figure}
This algorithm works on the principle of linear regression but from the end of the metric time series. Fig.~\ref{fig_overall_lbr} shows the overall workflow of this algorithm. Metric time series data belonging to a VM  is passed by the \textit{Data Pre-Processing} module to the algorithm's \textit{Trend Lines Fitting} module. In \textit{Trend Lines Fitting} module, a sequence of observations $x_{t - L:t}$ of length $L +1$ equal to the minimum window size $W_{min}$ starting from the end of the metric time series are selected and linear regression is applied on them to fit a line. The R-squared score is calculated to determine how close the data is to the fitted line. The fitted line model and R-squared score are further passed to the \textit{Analysis} module, where it compares the calculated R-squared score with the defined minimum R-squared score $R_{min}$  and uses the fitted line model to calculate the time to reach the defined threshold $U$. If the calculated R-squared score is above the defined $R_{min}$ and time to reach threshold is within the defined critical time then that sequence of observations is marked as anomalous, otherwise normal. This procedure is repeated by adding the next previous observation in the data until all the observations are used or reached a maximum window size $W_{max}$.  If multiple lines for different window sizes comply with the criteria then, the algorithm picks the line with the longest window size and all the observations in that window are marked as anomalous. The whole algorithm's pseudo code is shown in the  algorithm~\ref{alg:lbr}. As the algorithm applies linear regression but from the end of metric time series that is why it is named Linear Backward Regression (LBR).

%Fig.~\ref{fig_lbr} shows the memory utilization of an example VM with multiple fitted-lines of different window sizes $\{W_1, W_2, ..., W_n\}$  satisfying the criteria and $ W_{min} \leq W_1, W_2, ..., W_n \leq W_{max} $. The algorithm picks up the longest window $W_n$ and flags the whole window as an anomalous.
%\begin{figure}[htbp]
%\centerline{\includegraphics[width=1\linewidth]{figures/lbr.png}}
%\caption{Example of Linear Backward Regression (LBR) algorithm where the %longest window $W_n$ is marked as anomalous. }
%\label{fig_lbr}
%\end{figure}

\begin{algorithm}
\caption{Linear Backward Regression algorithm}\label{alg:lbr}
\begin{algorithmic}[1]
\Procedure{LBR}{$\textit{x=inputTs}$} \Comment{Input data: \textit{inputTs}}
\State ${a}\gets \text{[0] }$ \Comment{anomalous empty array of size \textit{x}}
\State ${W_{min}, W_{max}, R_{min}, C}\gets \text{User-defined values }$
\State ${U}\gets \text{100 }$\Comment{threshold utilization}
\State ${i}\gets {W_{min}}$ \Comment{i is loop variable}

\While{$i\leq length(x) \text{ and } i\leq W_{max} $}
\State $\textit{ts}\gets x[length(x) - i : length(x)]$
\State $model, r2 \gets \textit{\textbf{LinearRegression}(ts)}$
\State ${time}\gets \text{get time to } \textit{U} \text{ using }\textit{model}$

\If{$r2\geq R_{min} \text{ and } time\leq C $}
\State $a[length(x) - i : length(x)] \gets 1$
\EndIf
\State $i\gets i+1$
\EndWhile
\State \textbf{return} $a$\Comment{anomalous data represented by 1}
\EndProcedure
\end{algorithmic}
\end{algorithm} 

\subsection{Linear Backward Regression based on Change Points Detection (LBRCPD)}

The LBR algorithm described in the last subsection selects each observation one by one after the initial minimum window size for analyzing different window sizes. Thus, its computation time can be high for large time series data. Therefore a modification to LBR as this new algorithm is introduced, to overcome the problem of the computation time.  The idea behind this modification is to instead of going over each observation, we directly jump to the change point observation for moving from one window size to another while fitting the line. Change points represent the observations where a significant change has occurred to the last normal observations. 

In this algorithm, firstly, on the whole dataset the change points $P = \{P_1, P_2, ..., P_k\}$, where $k \leq n-1$,  are detected. This is done by first taking the first-order difference of the input data. Then, taking their absolute values and calculating Z-scores for them. Further, the indexes are returned for observations whose Z-scores are greater than the defined threshold (3 times the standard deviation). The method's pseudocode is shown in the  algorithm's~\ref{alg:lbrcpd} \textit{CPD procedure}.  This method detects the observations which deviate highly from the normal pattern of the data referred as change points and these points are used by the algorithm to select a different window size. By default, two change points one at the beginning and other at the end of time series data are added. Also, the required minimum difference between the two change points is set as part of this algorithm so that we don't get many change points.

This algorithm has a similar design as LBR and works as follows. In \textit{Trend Lines Fitting} module, a sequence of observations $x_{t - L:t}$ between the last two change points $P_k \text{ and } P_{k-1}$ (where $L$ is equal to the difference between the last two change points)  starting from the end of the metric time series are taken and a line is fitted on them using the linear regression. The R-squared score is calculated to determine the approximation of the data to the fitted line. As in LBR, the fitted line model, and R-squared score are passed further to the \textit{Analysis} module, where it compares the calculated R-squared score with the defined minimum R-squared score $R_{min}$  and uses the fitted line model to calculate the time to reach the defined threshold. If the calculated R-squared score is above the defined minimum R-squared score and time to reach threshold is within the defined critical time then that sequence of observations is marked as anomalous, otherwise normal. This procedure is further repeated by analyzing the observations between the last change point $P_k$ and the next previous change point $P_{k-2}$ until all the change points are used. Similarly, like LBR, if multiple lines for different window sizes comply with the criteria then, the algorithm picks the line with the longest window size and all the observations in that window are marked as anomalous. The algorithm's pseudocode is shown in the  algorithm~\ref{alg:lbrcpd}. As the algorithm applies change points detection and LBR, that is why it is named Linear Backward Regression based on Change Points Detection (LBRCPD). This method reduces the computation time significantly with almost the same accuracy.

\begin{algorithm}
\caption{LBR based on Change Points algorithm}\label{alg:lbrcpd}
\begin{algorithmic}[1]\label{alg:cpd}
\Procedure{CPD}{$\textit{x=inputTs}$}\Comment{Change-points on $\textit{inputTs}$}
\State $\textit{threshold}\gets 3$
\State $\textit{diffTs} \gets \text{first order difference of }\textit{x}$
\State $\textit{absDiffTs}\gets \text{absolute values of }\textit{diffTs}$
\State $\textit{zScores}\gets \text{calculate z-scores of }\textit{absDiffTs}$
\State $\textit{cpdIndexes}\gets \text{indexes of }\textit{(zScores} > \textit{threshold)}$
\State \textbf{return} $\textit{cpdIndexes}$\Comment{The change-points indexes}
\EndProcedure

\Procedure{LBRCPD}{$\textit{x=inputTs}$}\Comment{Input data: \textit{inputTs}}
\State ${a}\gets \text{[0] }$ \Comment{anomalous empty array of size \textit{x}}
\State ${R_{min}, C, U}\gets \text{User-defined values}$
\State ${P}\gets \textit{\textbf{CPD}(x)}$ \Comment{get change point indexes}
\While{$i\leq length(P)$} \Comment{i=1 is a loop variable}
\State $\textit{ts}\gets x[P[length(P)-i ] : P[length(P)]]$
\State $model, r2 \gets \textit{\textbf{LinearRegression}(ts)}$
\State ${time}\gets \text{get time to } \textit{U} \text{ using }\textit{model}$
\If{$r2\geq R_{min} \text{ and } time\leq C $}
\State $a[P[length(P)-i ] : P[length(P)]] \gets 1$
\EndIf
\State $i\gets i+1$
\EndWhile
\State \textbf{return} $a$\Comment{anomalous data represented by 1}
\EndProcedure
\end{algorithmic}
\end{algorithm}

\subsection{Precog}
The previous two algorithms introduced do not use historic data information for making smart decisions. For example, LBR and LBRCPD will capture a trend in the new data if its duration is greater than the defined minimum trend duration and satisfies the other criteria of minimum R2 and time to reach the threshold. However, if a similar trend exists in the historic data and reflects the normal memory usage of the application then, such a trend should be ignored. This will reduce the number of false positives. Therefore, to make the detection process smarter, the Precog algorithm is developed. 
\begin{figure}[htbp]
\vspace{-2mm}
\centerline{\includegraphics[width=1\linewidth]{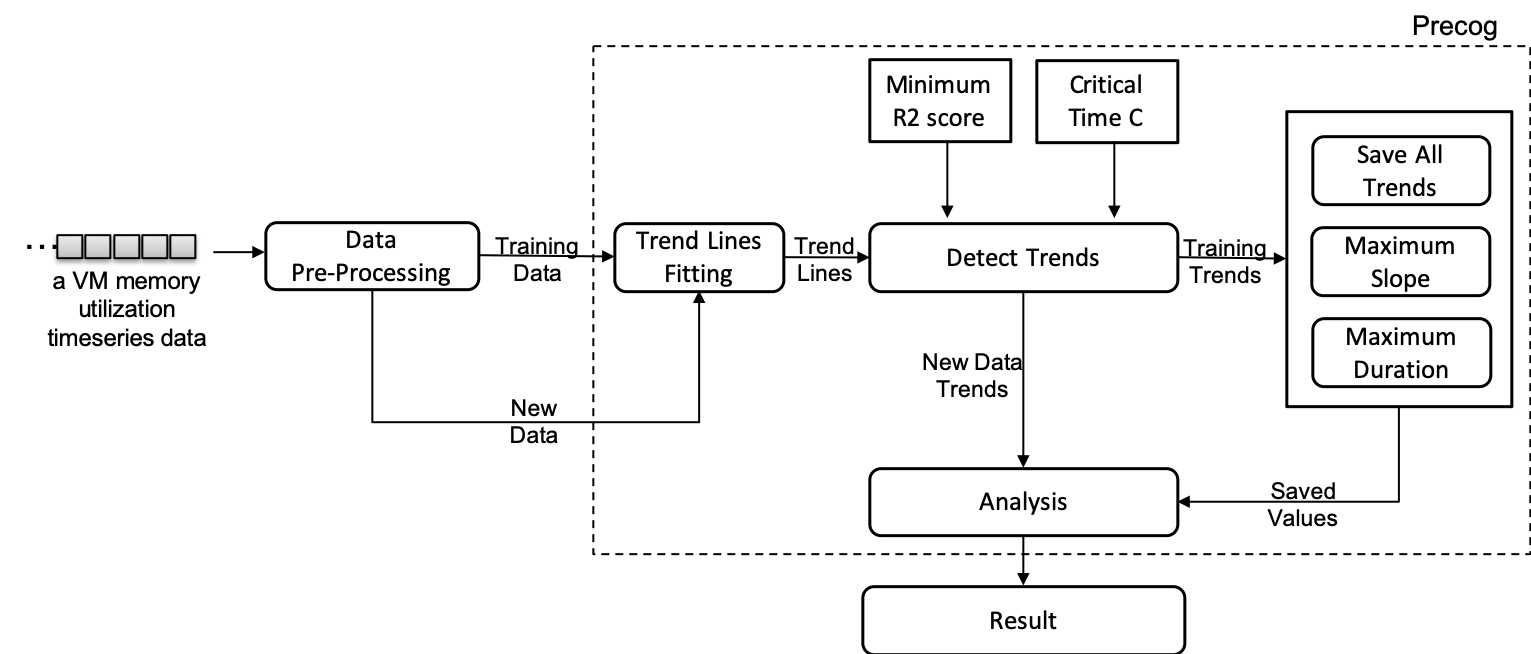}}
\caption{Overall workflow of Precog algorithm. }
\label{fig_precog}
\vspace{-4mm}
\end{figure}

Fig.~\ref{fig_precog} shows the overall workflow of the 
Precog algorithm. The whole procedure consists of two phases: offline training and online detection. \textit{Data Pre-Processing} and \textit{Trend Lines Fitting} modules are shared by both the phases. 

Metric time series data belonging to a VM  is passed by the \textit{Data Pre-Processing} module to the algorithm's \textit{Trend Lines Fitting} module. In \textit{Trend Lines Fitting} module, firstly, on the whole dataset whether it is the historic data or new data, the change points $P = \{P_1, P_2, ..., P_k\}$, where $k \leq n-1$,  are detected.

In the offline training phase, \textit{Trend Lines Fitting} module further selects a sequence of observations $x_{t - L:t}$ between the two change points: one fixed $P_1$ and other variable $P_{r}$ where $r \leq k$  and a line is fitted on them using the linear regression. The R-squared score, size of the window called as duration, time to reach threshold and slope of line are calculated. This procedure is repeated with keeping the fixed change point the same and varying the other for all other change points. Out of all the fitted lines, the best-fitted line based on the largest duration and highest slope is selected for the fixed change point. If this best-fitted lines' time to reach threshold falls below the critical time then its slope and duration are saved as historic trends. This above procedure is again repeated by changing the fixed change point to all the other change points. At the end of this whole procedure, we get for each change point, a best-fitted trend if it exists. Amongst the captured trends, maximum duration and the maximum slope of the trends are also calculated and saved. This training procedure can be conducted routinely, e.g., once per day or week. The method's pseudocode is shown in the  algorithm's~\ref{alg:precog} \textit{Train procedure}.

%As in LBRCPD, the best-fitted line's R-squared score is compared with the defined minimum R-squared score $R_{min}$ and uses the best-fitted line model to calculate the time to reach the defined threshold. If the calculated R-squared score is above the defined minimum R-squared score and time to reach threshold is within the defined critical time then  

In the Online Detection phase, for a new set of observations $\{x_k, x_k+1, x_k+2, ...,x_k+t-1 x_k+t\}$ from time $k$ to $t$ where $t - k \geq P_{min}$ belonging to a VM  after pre-processing is fed into the \textit{Trend Lines Fitting} module.  In \textit{Trend Lines Fitting} module, the change points are detected. Similar to LBRCPD, a sequence of observations $x_{t - L:t}$ between the last two change points starting from the end of the time series are taken and a line is fitted on them using the linear regression. The R-squared score, slope, duration and time to reach threshold of the fitted line is calculated. If its slope and duration are greater than the saved maximum counter parts then that window is marked anomalous. Otherwise, the values are compared against all the found training trends and if fitted-line's slope and duration are found to be greater than any of the saved trend then, again that window will be marked as anomalous. This procedure is further repeated by analyzing the observations between the last change point $P_k$ and the previous next change point until all the change points are used. The algorithm's pseudo code showing the training and test method are shown in the  algorithm~\ref{alg:precog}.

%New trends and training trends can further be combined if possible to form a bigger trend for better analysis but that part is not yet implemented as part of the algorithm.

\begin{algorithm}
\caption{Precog Algorithm}\label{alg:precog}
\begin{algorithmic}[1]
\Procedure{Train}{$\textit{x=inputTs}$}\Comment{Training on historic  data}
\State ${R_{min}}\gets \text{User-defined value }$
\State ${P}\gets \textit{\textbf{CPD}(x)}$ \Comment{get change point indexes}
\State ${p1}\gets {0}$ \Comment{p1 is a loop variable}
\While{$p1\leq length(P)$}
\State $p2 \gets p1$
\State ${D_{b}, S_{b}, T_{b}}\gets 0$ \Comment{best local values}
\While{$p2\leq length(P)$}
\State $\textit{ts}\gets x[P[p1] : P[p2]]$
\State $time, r2, d, s \gets \textit{\textbf{LinearRegression}(ts)}$
\If{$r2\geq R_{min} \text{ and } d\geq D_{b} \text{ and } s\geq S_{b}$}
\State  \textbf{Update}(${D_{b}, S_{b}, T_{b}}$) \Comment{best local values}
\EndIf
\State $p2\gets p2+1$
\EndWhile
\If{$T_{b}\leq C$}
\If{$D_{b}\geq D_{max} \text{ and } S_{b}\geq S_{max}$}
\State  \textbf{Update}(${D_{max}, S_{max}}$) \Comment{update values}
\EndIf
\State \textbf{saveTrend}($D_{b}, S_{b}), \textbf{ save}({D_{max}}, S_{max}$)
\EndIf
\State $p1\gets p1+1$
\EndWhile
\EndProcedure
\Procedure{Test}{$\textit{x=inputTs}$}
\State ${a}\gets \text{[0] }$ \Comment{anomalous empty array of size \textit{inputTs}}
\State ${P}\gets \textit{\textbf{CPD}(x)}$ \Comment{get change point indexes}
\State ${len}\gets \text{length(P)}$ \Comment{length of change point indexes}
\While{$i\leq len$} \Comment{i=1 is a loop variable}
\State $\textit{ts}\gets x[P[len-i ] : P[len]]$
\State $time, r2, d, s \gets \textit{\textbf{LinearRegression}(ts)}$

\State ${ D_{max}, S_{max}, Trends}\gets \text{get saved values }$
\If{$time\leq C \text{ and } r2\geq R_{min} $}
\If{$ s\geq S_{max} \text{ and } d\geq D_{max}$}
\State $a[P[len-i ] : P[len]] \gets 1$
\Else
\State $\textbf{For Each } t \text{ in } Trends$
\If{$ s\geq S_{t} \text{ and } d\geq D_{t}$}
\State $a[P[len-i ] : P[len]] \gets 1$
\EndIf
\EndIf
\EndIf
\State $i\gets i+1$
\EndWhile
\State \textbf{return} $a$\Comment{anomalous data represented by 1}
\EndProcedure
\end{algorithmic}
\end{algorithm}

\subsection{Precog with Maximum Filtration (PrecogMF)}
It was found that after running the Precog on a sample dataset for evaluation, there were cases where there exists an increasing trend in the new data but the overall maximum of this trend is less than the training data. As a result, these trends were getting reported as anomalous but should not have been as the training data has seen a similar trend with a higher value observation hence this increasing trend is normal and can be ignored. 

Thus, a modification to Precog is introduced called Precog with Maximum Filtration to check if the reported anomalous window's maximum value is less than the one found in the training data. If it is found to be less, then the window is declared as normal otherwise, not. Thus further increasing the accuracy of the algorithm. 
\section{Evaluation}
\subsection{Experimental Settings}

To demonstrate the effectiveness of the developed algorithms, we conduct experiments on the real Cloud dataset provided by our partner industry Huawei which consists of manually labeled memory leak data from 60 VMs spanned over 5 days and each time series consists of an observation every minute. Out of these 60 VMs, 20 VMs had a memory leak. We have used F1-Score (denoted as F1) to evaluate
the performance of the algorithms: $F1 = \frac{2 \times Precision \times Recall}{Precision+Recall}$, where $Precision= \frac{TP}{TP+FP}$, $Recall= \frac{TP}{TP+FN}$. Evaluation tests have been executed on a machine with 4 physical cores (3.6 GHz Intel Core i7-4790 CPU) with hyperthreading enabled  and 16 GB of RAM. These conditions are similar to a typical cloud VM. Hyper-parameters set in our experiments common to all the algorithms are as follows. The minimum duration required for a trend is set to 6 hours. The maximum threshold $U$ is set to 100 and the defined critical time $C$ is set to 7 days. The smoothing window size is 1 hour and re-sampling time resolution was set to 5 minutes. Lastly, the minimum R-squared score $R2_{min}$ for a line to be recognized as a good fit is set to 0.8. Hyper-parameters other then the common ones which are confined to the individual algorithm are as follows. The starting minimum window size $W_{min}$ in LBR for analysis is set equal to the required minimum trend duration (6 hours) while the maximum window size $W_{max}$ is set to 7 days.  Algorithms using change points have a minimum distance between two change points $P_{min}$ equal to the required minimum trend duration (6 hours). For Precog and PrecogMF, 65\% of data was used for training and the rest for testing.

\subsection{Results}
In this subsection, we first present the detail results of the individual algorithm and then later compare their performance.
\begin{figure*}[htbp] 
\begin{subfigure}{.33\textwidth}
  \centering
  \includegraphics[width=0.97\linewidth]{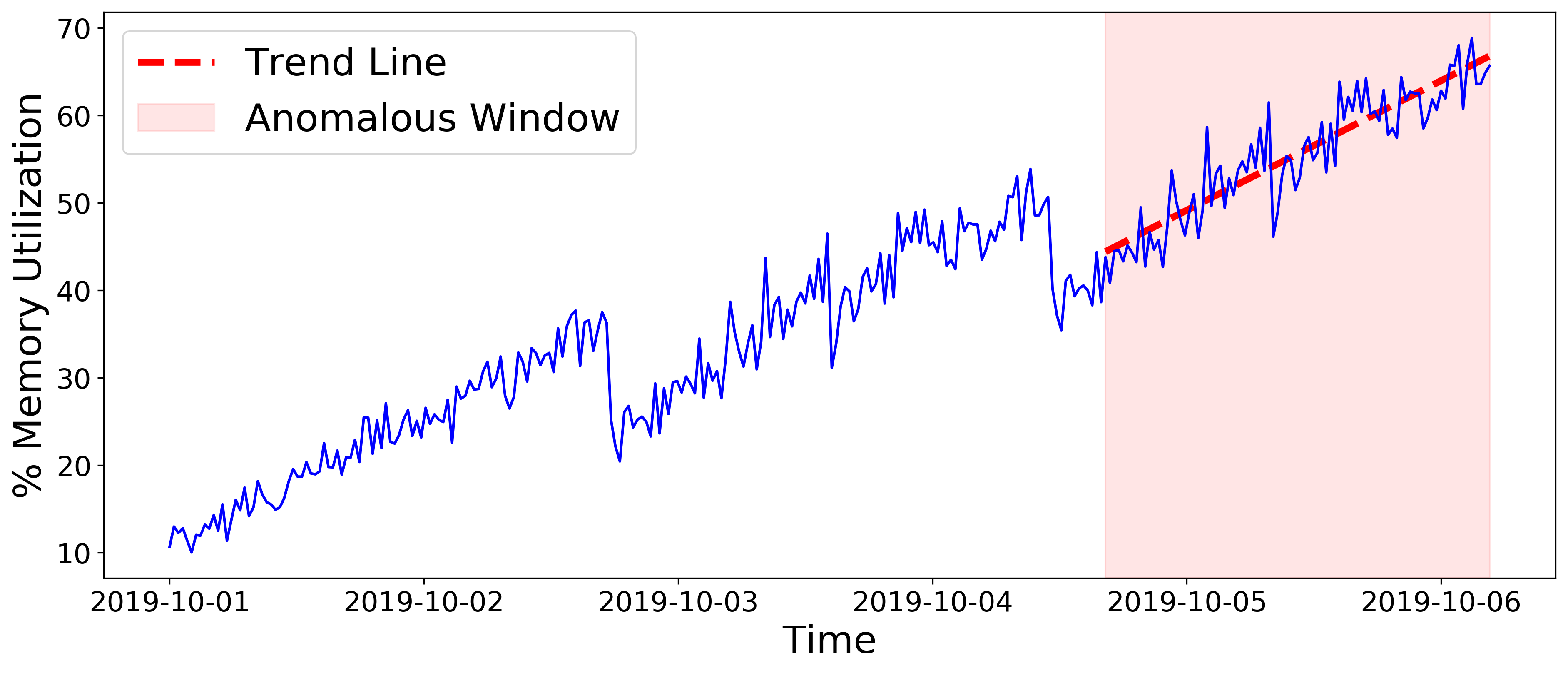}
\end{subfigure}%
\begin{subfigure}{0.33\textwidth}
  \centering
  \includegraphics[width=0.97\linewidth]{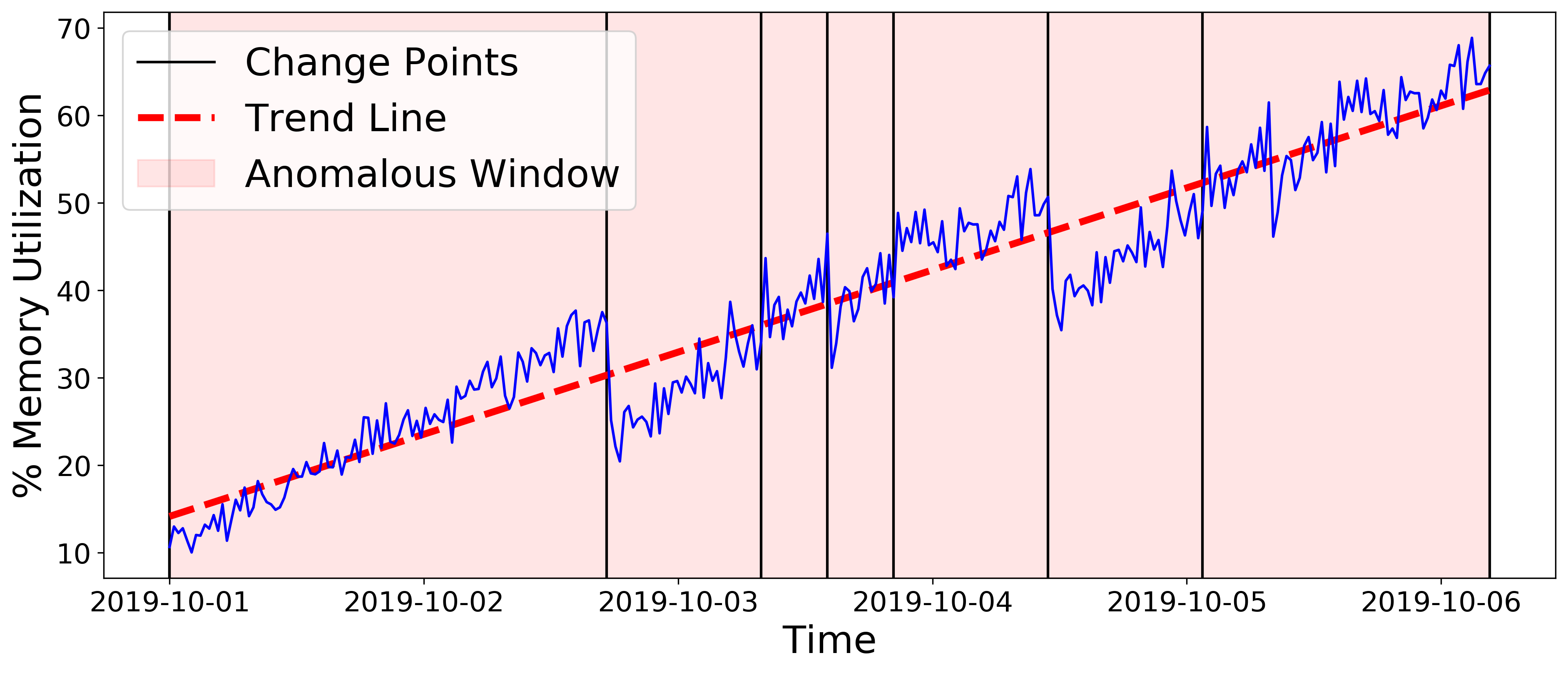}
\end{subfigure}
\begin{subfigure}{0.33\textwidth}
  \centering
  \includegraphics[width=0.97\linewidth]{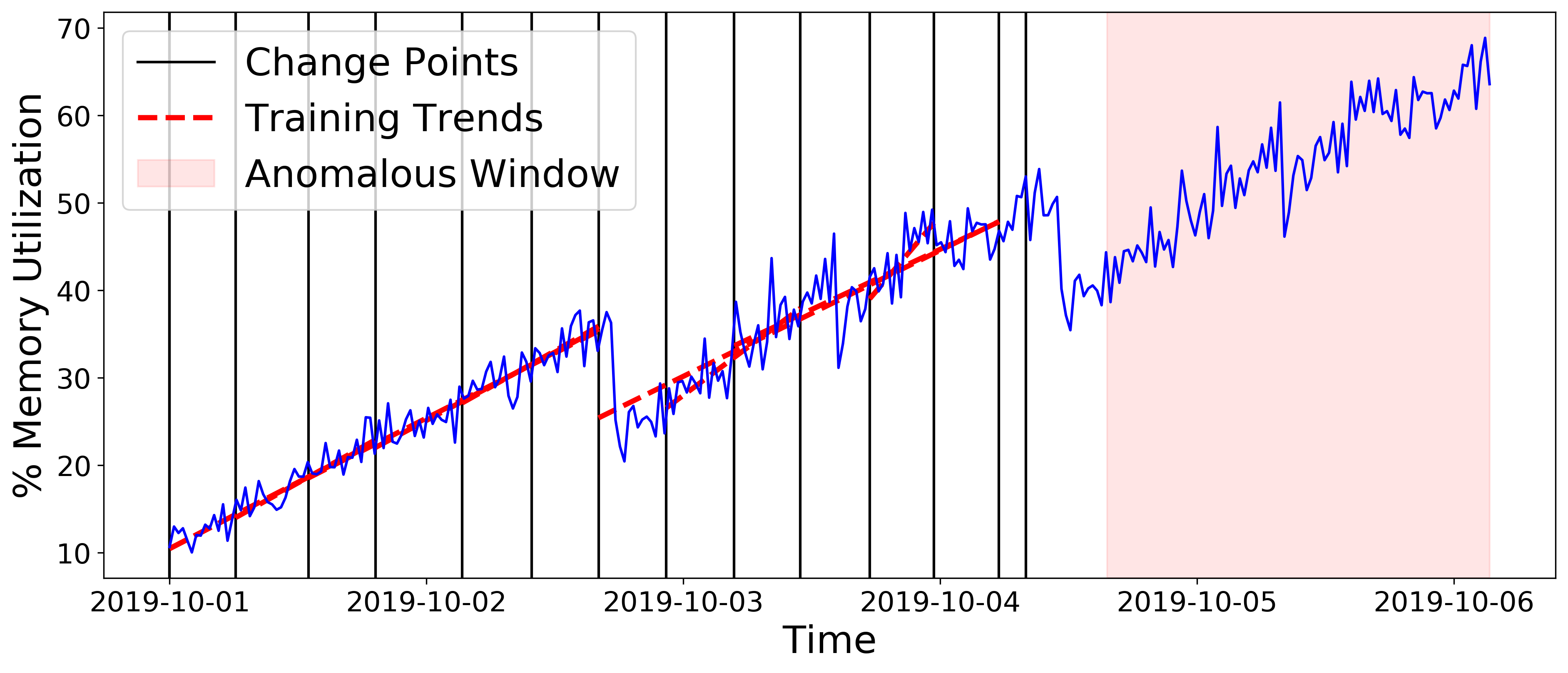}
\end{subfigure}
\begin{subfigure}{.33\textwidth}
  \centering
  \includegraphics[width=0.97\linewidth]{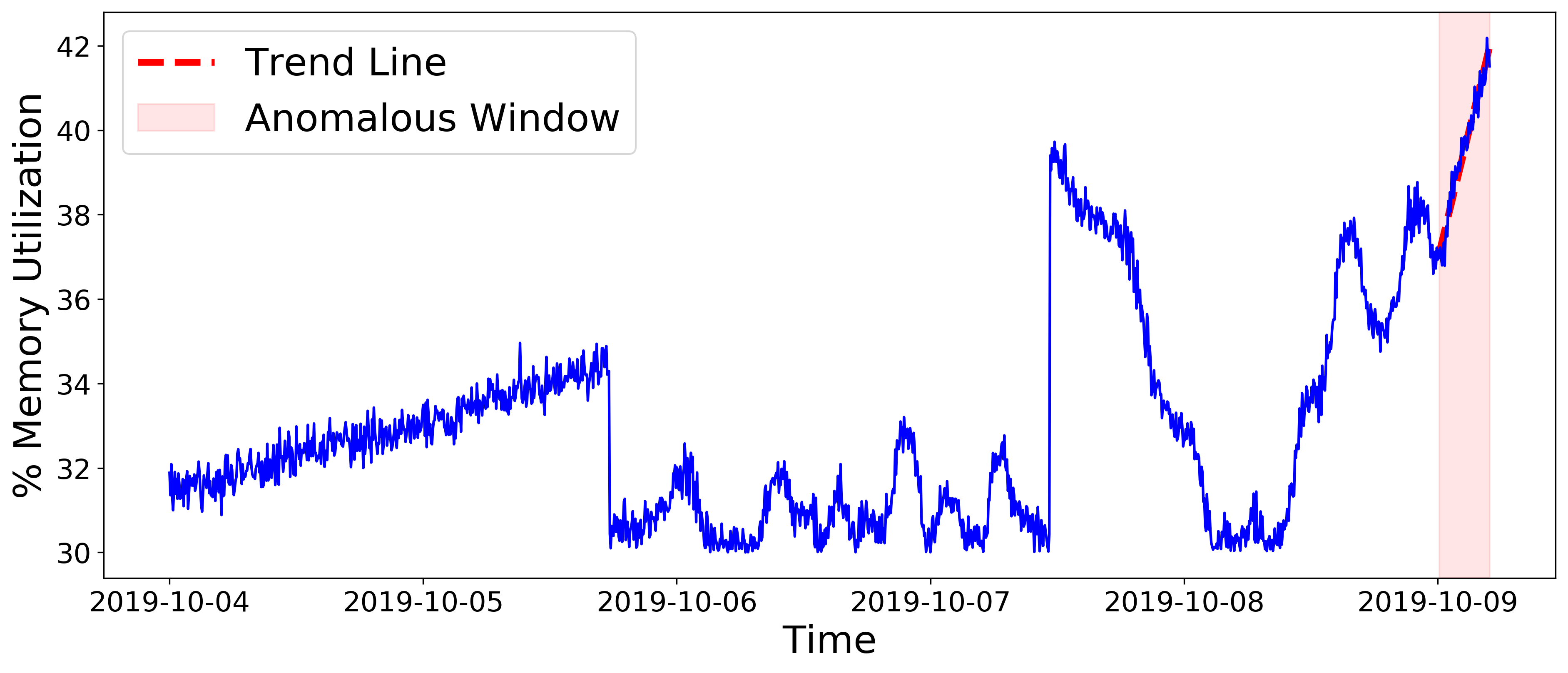}
  \captionof{figure}{LBR}
  \label{fig:lbr_perf}
\end{subfigure}%%
\begin{subfigure}{0.33\textwidth}
  \centering
  \includegraphics[width=0.97\linewidth]{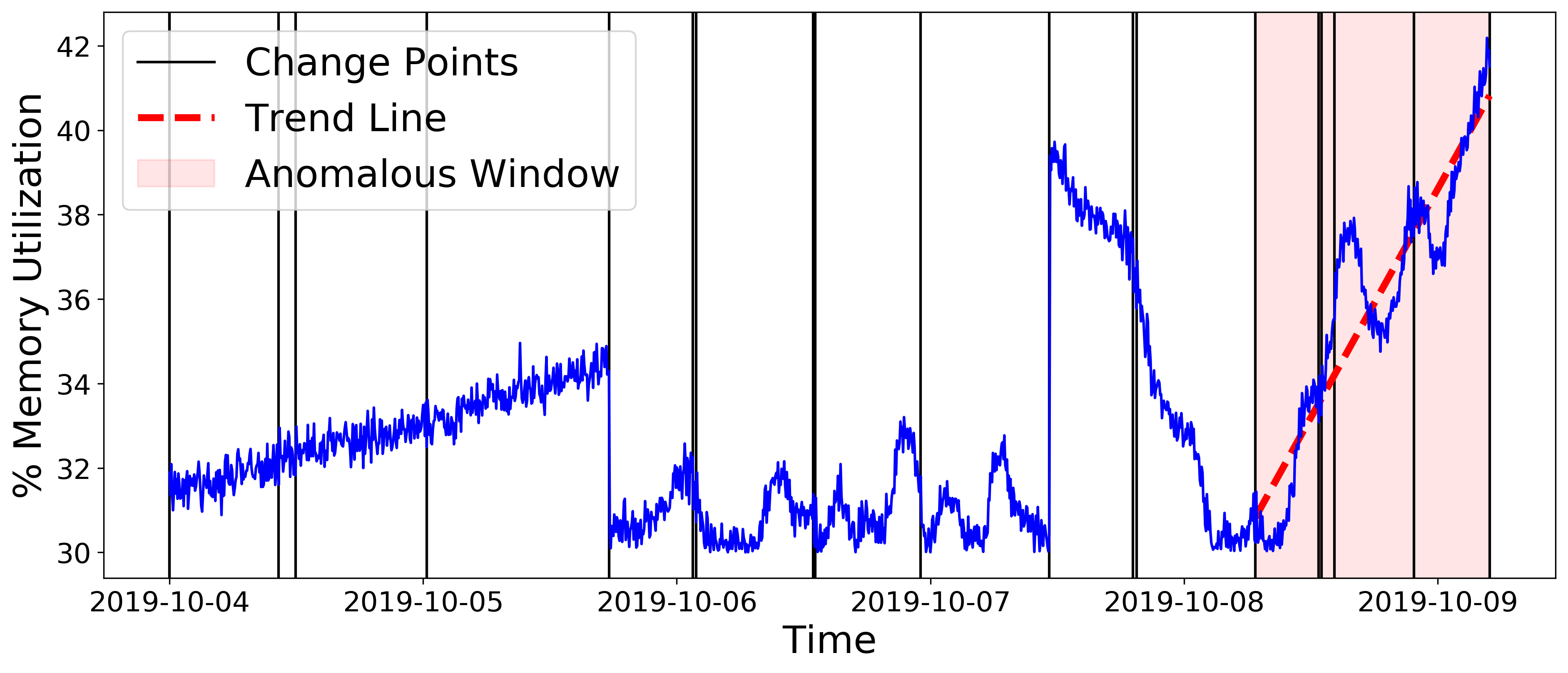}
  \captionof{figure}{LBRCPD}
  \label{fig:lbr_cpd_perf}  
\end{subfigure}
\begin{subfigure}{0.33\textwidth}
  \centering
  \includegraphics[width=0.97\linewidth]{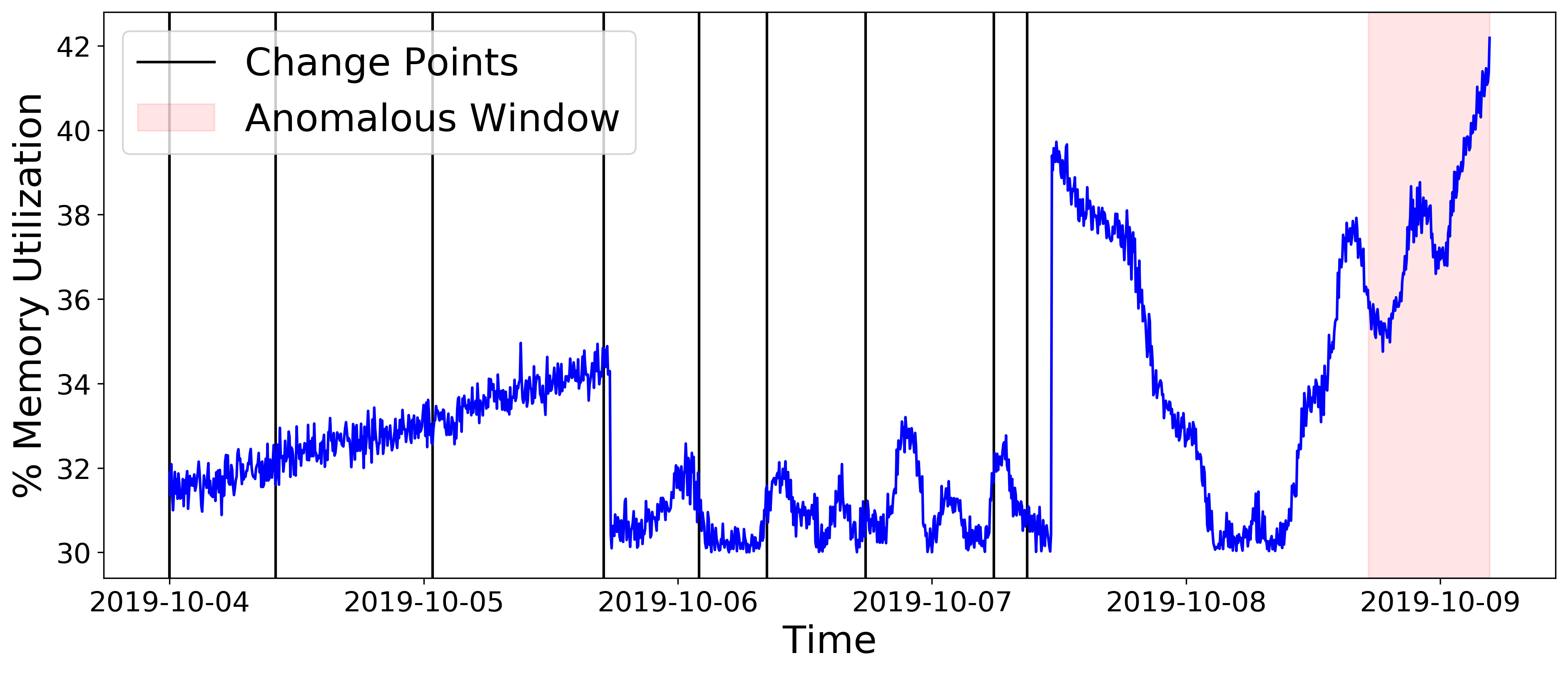}
  \captionof{figure}{Precog}
  \label{fig:precog_perf}  
\end{subfigure}
 \caption{Algorithms result when applied on the two VM's memory utilization data having a memory leak.}
  \label{algorithms_test_results}
\end{figure*}

Fig.~\ref{algorithms_test_results} shows the result of the three algorithms: LBR, LBRCPD, and Precog on two VMs (shown by two rows) memory utilization data having a memory leak. All the three algorithms were able to detect the memory leak on these VMs however the size of the anomalous window (marked by red) varies from each other. For LBR and LBRCPD there is no separate training and detection part, therefore, for them, the whole data was passed for detection while in the case of Precog, 65\% of data was used for the training and rest as testing set. 

In Fig.~\ref{fig:lbr_perf}, one can see that the LBR algorithm stopped its analysis once the fitting approximation of the line starts to degrade and reaches less than the defined minimum R-squared for both the cases. In the first case despite having a global increasing trend, the algorithm detected the local one. Though in these two cases it has worked perfectly fine, it may not account for global trend in the cases where there is a short burst of increment or decrement in memory usage in between the increasing memory leak trend.

In the case of LBRCPD, change points are also shown in Fig.~\ref{fig:lbr_cpd_perf} along with the trend line and anomalous window for both the cases.  LBRCPD captured the global increasing trend of the data due to the window size selection based on the change points. This jumping from one change point to another allows this algorithm not to stuck in the local trend as was in the case with LBR. Though this captures the global memory leak trend but misses up the local trend which can be troublesome if the local trend is approaching towards maximum utilization. One can think of making a hybrid of LBR and LBRCPD to capture both the local and global trends. 

The results of the Precog algorithm are shown in Fig.~\ref{fig:precog_perf} along with the training trends, change points and the anomalous window. For the first case, one can see that the algorithm detected two major big training trends and reported most of the test set as anomalous. The test set in it follows a similar trend as captured during the training but with the higher memory utilization, hence it is reported. For the second case, no trend was detected in the complete training data but still, the algorithm was able to detect an anomalous window in the test data due to the usage of a similar algorithm as LBRCPD on the test data set and therefore found an increasing memory leak trend.
\begin{figure*}[htbp] 
  \centering
  \begin{subfigure}{.33\textwidth}
  \centering
  \includegraphics[width=0.97\linewidth]{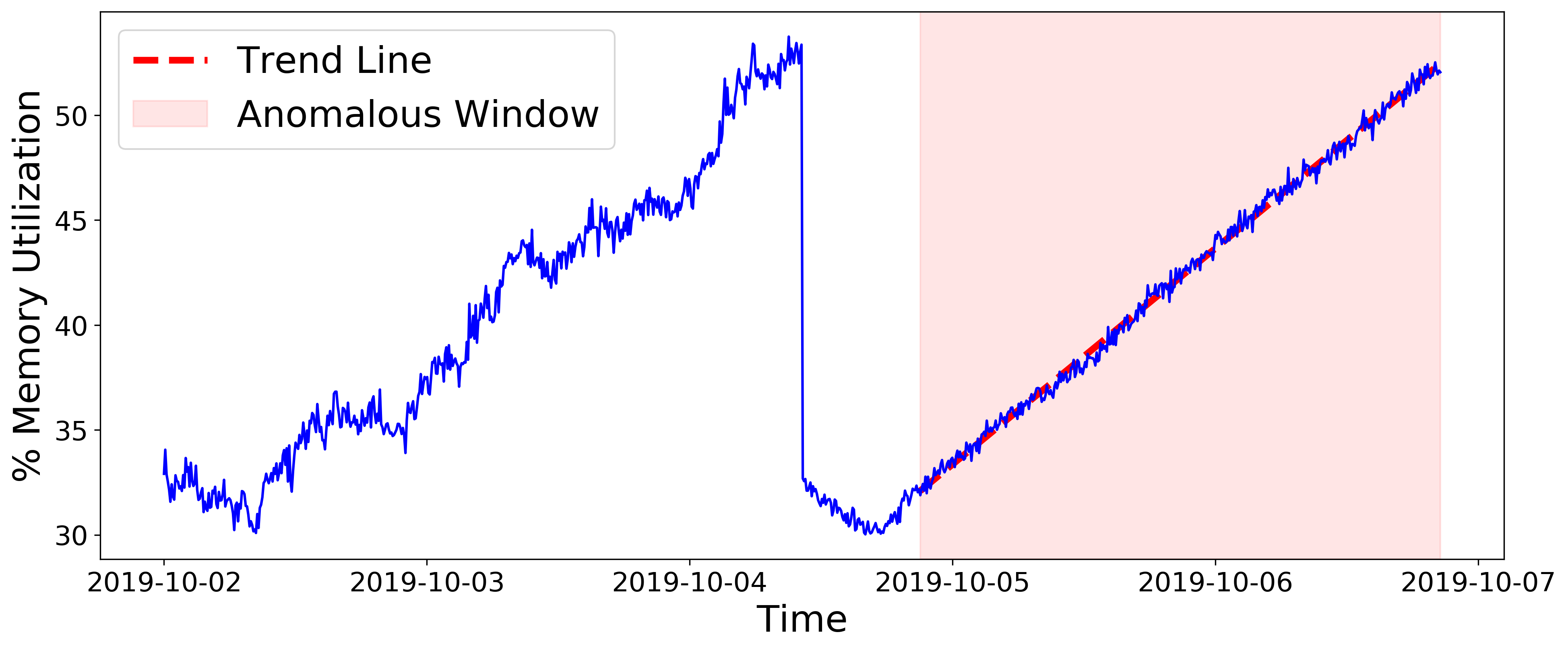}
    \captionof{figure}{LBR}
  \label{fig:lbr_perf_not_case}
\end{subfigure}%
\begin{subfigure}{0.33\textwidth}
  \centering
  \includegraphics[width=0.97\linewidth]{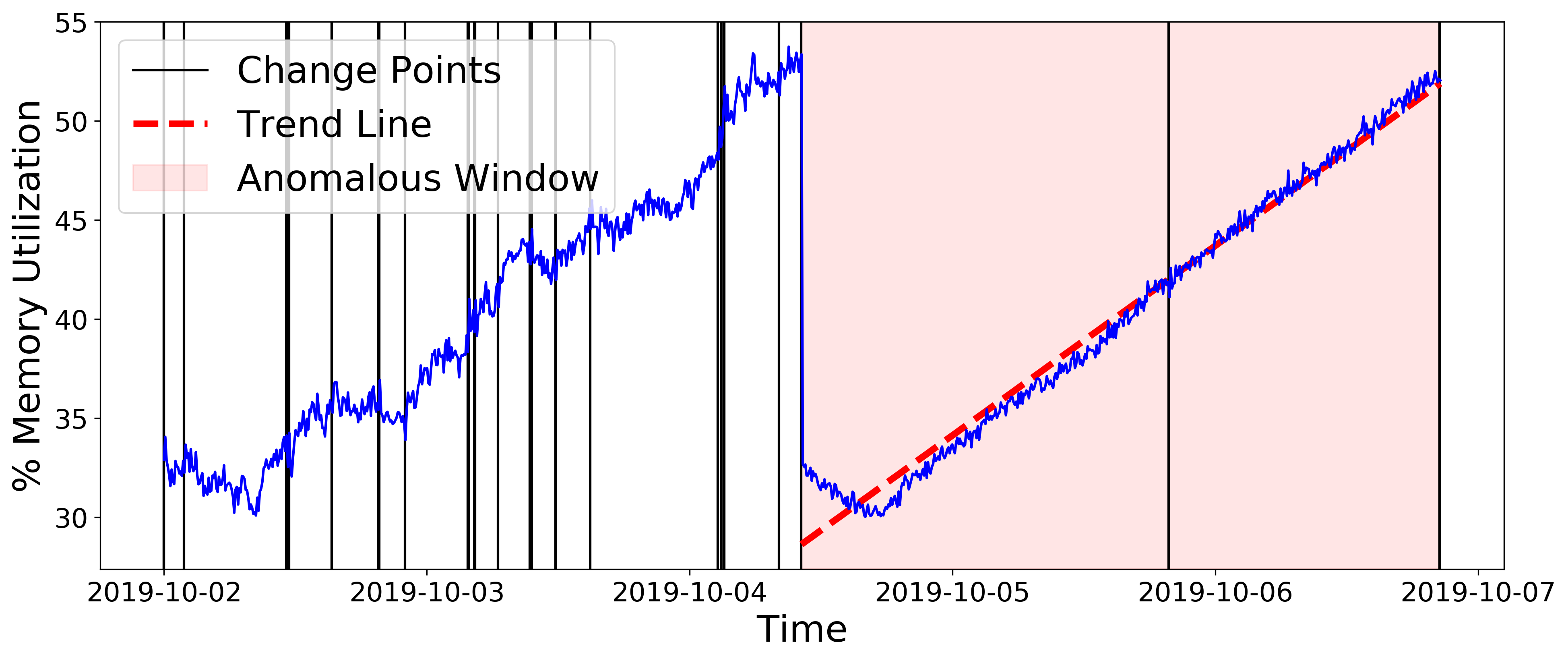}
    \captionof{figure}{LBRCPD}
  \label{fig:lbr_cpd_perf_not_case}  
\end{subfigure}
\begin{subfigure}{0.33\textwidth}
  \centering
  \includegraphics[width=0.97\linewidth]{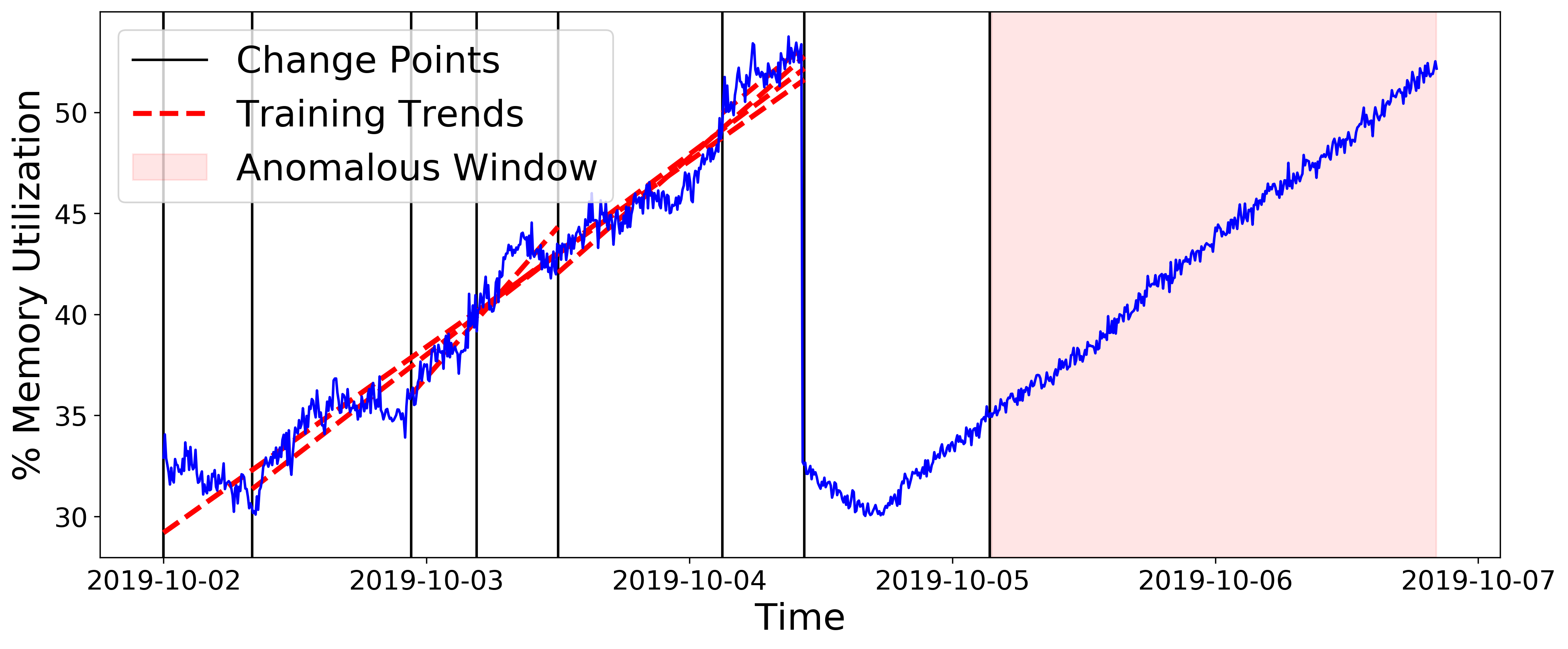}
    \captionof{figure}{Precog}
  \label{fig:precog_perf_not_case}  
\end{subfigure}
 \caption{3 algorithms wrongly show anomalous window on the VM's memory utilization data not having a memory leak.}
  \label{result_on_vm_not_having_memory_leak}
       \vspace{-4mm}
\end{figure*}

\begin{figure}[htbp] 
  \centering
  \includegraphics[width=0.97\linewidth]{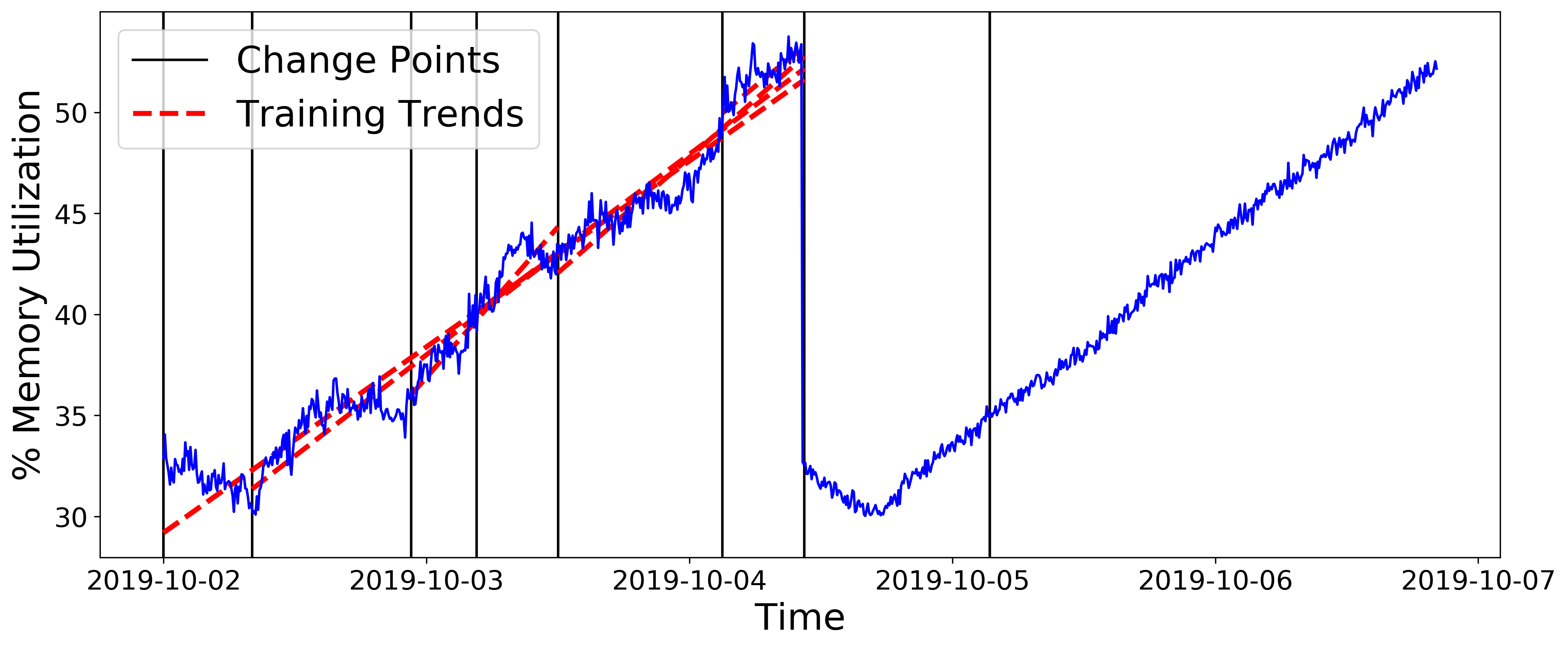}

  \caption{PrecogMF does not shows anomalous window on a VM's memory utilization data not having a memory leak.}

  \label{fig:Precog_mf_result_on_vm_not_having_memory_leak}
\end{figure}

Fig.~\ref{result_on_vm_not_having_memory_leak} shows the result of the three algorithms: LBR, LBRCPD, and Precog when applied on a VM that does not have a memory leak but its memory utilization was steadily increasing. It can be seen that all the three algorithms show an anomalous window predicting a memory leak wrongly on that VM. However, when PrecogMF was applied to the same data, it didn't report a memory leak on it because of the maximum filtration part added into it. As the new data's maximum value is less than the training data and hence can be deemed normal. This is shown in Fig~\ref{fig:Precog_mf_result_on_vm_not_having_memory_leak}. It is also to be noted that, if the new data's maximum goes beyond the maximum in the training data then it will be regarded as a memory leak case. 

We further have conducted the algorithms performance evaluation test by recording the F1 score on the dataset provided by Huawei along with the time required to complete the evaluation. It should be noted that the time calculation for Precog and PrecogMF includes the training time as well. Results are summarized in the Table~\ref{performance_results_algorithms}. One can see that the time required for computation using LBR is highest. It is reduced significantly in the other algorithms after the usage of change points. The accuracy hike from LBR to Precog is almost 10\% but the compute time is reduced significantly by 80\%, which is beneficial for online real-time discovery of memory leaks in the cloud-based infrastructure. After the usage of PrecogMF, the accuracy gain from LBR is almost 30\%. 

\begin{table}[ht]
\caption{F1-score and time taken by algorithms for analysis}
\centering
\begin{tabular}{|p{0.15\linewidth}|p{0.15\linewidth}|p{0.2\linewidth}|p{0.2\linewidth}|}
\hline
\textbf{Algorithm}&\textbf{F1-Score}&\textbf{Overall Time (in seconds)}&\textbf{Time per test (in seconds)}\\
\hline
LBR& 0.568 & 97.964& 1.63\\
\hline
LBRCPD & 0.625 & 23.922 & 0.398\\
\hline
Precog & 0.666 &  19.508 &  0.325\\
\hline
PrecogMF & 0.857 & 19.5177 & 0.3252\\
\hline
\end{tabular}
\label{performance_results_algorithms}
\vspace{-4mm}
\end{table}

\section{Discussion}
To summarize, in most of the experimental scenarios, an application having a memory leak shows either a steadily linearly increasing or sawtooth memory utilization pattern. The 4 different algorithms proposed as part of this work can be used depending upon the use case. LBR though being a compute-intensive does provide a stable and simple way for detection of linearly increasing memory leaks. It can be used on the systems when there is no historic memory usage data of the applications and one wants to detect the memory leaks in a not so much complex way. On the other hand, LBRCPD offers a less compute-intensive way of discovering a memory leak with better or almost the same accuracy as LBR. However, choosing which change detection method to use and tuning of the other parameters like the minimum distance between two change points can be challenging. But one can afford the tuning of the hyper-parameters as it is less compute-intensive. Precog being the smartest of all the algorithms presented uses the historic data to estimate the parameters and hence would be much suited for the cases where there are already some available memory usage data of the applications. Lastly, PrecogMF offers a variant on top of the Precog to detect the normal increasing memory usage pattern of an application and reduce false positives. 

In general, most of the algorithms were able to detect the memory leak, however, the size of the anomalous window and the trend line reported by them differs from each other. One can use the size of the window to determine the confidence of the memory leak, like if it is big then it has higher confidence of a memory leak.  In the cases where the application is using the memory steadily for normal processing, the three algorithms LBR, LBRCPD, and Precog wrongly predict the memory leak as shown in Fig.~\ref{result_on_vm_not_having_memory_leak}. PrecogMF does not detect these normal cases as memory leak cases. 

Further, evaluation tests reveal that PrecogMF shows the highest accuracy of 85\% with 325 milliseconds required for evaluation per time series data. This gives the most suitable use case for the cloud-based infrastructure where multiple VMs are to be tested for memory leaks without the knowledge of the applications deployed on them and in real-time. One can also think of using the hybrid of all the introduced algorithms. For example, using the LBR or LBRCPD at the start of the deployment of the application and once there is some data then shifting to PrecogMF. In the background, LBRCPD can always run to identify the global trend to further increase the confidence of the discovery. 

\section{Conclusion}
Memory leak detection has been a research topic for more than a decade. Many approaches have been proposed to detect memory leaks, with most of them looking at the internals of the application or the object's allocation and deallocation. The algorithms for memory leak detection presented in the current work are most relevant for the cloud-based infrastructure where cloud administrator does not have access to the source code or know about the internals of the deployed applications. This work presented two novel algorithms: LBR and Precog and, their two variants: LBRCPD and PrecogMF for solving this challenge. The performance evaluation results showed that the accuracy increased by 30\% when using PrecogMF instead of LBR and reduced the overall time for analysis significantly by 80\%. The paper also presents the different memory leak patterns found in the applications written in different programming languages. It was found that in general an application having a memory leak exhibits either a steadily linearly increasing or sawtooth memory utilization pattern. 

Prospective directions of future work include developing online learning-based approaches for detection and as well using other metrics like CPU, network and storage utilization for further enhancing the accuracy of the algorithms and providing higher confidence in the detection results.
\section{Acknowledgements}
This work was supported by the funding of the German Federal Ministry of Education and Research (BMBF) in the scope of the Software Campus program. We also thank the Huawei Munich Research Center for providing the dataset which leads to the smooth conduct of this research work. 

\bibliographystyle{IEEEtran}  
\bibliography{bib}
\end{document}